\documentstyle[aps,graphics,epsfig,multicol]{revtex}

\def\PLB{{\em Phys. Lett.}  B}
\def\PHR{\em Phys. Rep.}
\def\PRL{\em Phys. Rev. Lett.}

\def\PRC{{\em Phys. Rev.} C}

\begin{document}


\vspace{5mm}
 
\begin{center}
{\bf \Large {Refractive elastic scattering of carbon and oxygen nuclei:
The mean field analysis and Airy structures}}\\
\vspace{4mm}
{\large  S. Szilner$^{a,b}$, M.P. Nicoli$^{b}$, Z. Basrak$^{a}$, 
R.M. Freeman$^{b}$, F. Haas$^{b}$,    
A. Morsad$^{c}$, M. E. Brandan$^{d}$, and G. R. Satchler$^{e}$ } \\

\vspace{4mm}
{\normalsize \it
$^a$Ru{d\llap{\raise 1.22ex\hbox
  {\vrule height 0.09ex width 0.2em}}\rlap{\raise 1.22ex\hbox
  {\vrule height 0.09ex width 0.06em}}}er
   Bo\v{s}kovi\'{c} Institute, HR-10$\,$002 Zagreb, Croatia \\
$^b$Institut de Recherches Subatomiques, 
CNRS-IN2P3/ULP,
Strasbourg, France \\
$^c$Facult\'e des Sciences Ben M'Sik, Universit\'e Hassan II, 
Casablanca, Morocco \\
$^d$ Instituto de Fisica, UNAM, Mexico, Mexico \\
$^e$ University of Tennessee, Knoxville and Oak Ridge National Laboratory, USA
}
\end{center}

 
\begin{abstract}

The experimental data on the $^{16}$O$+^{12}$C and $^{18}$O$+^{12}$C elastic
scatterings and their optical model analysis are presented.
Detailed and complete elastic angular distributions
have been measured at the Strasbourg Vivitron accelerator at several energies
covering the energy range between 5 and 10 MeV per nucleon.
The elastic scattering angular distributions show the usual
diffraction pattern and also, at larger angles, refractive effects
in the form of nuclear rainbow and associated Airy structures.
The optical model analysis unambiguously shows the evolution of the
refractive scattering pattern.
The observed structure, namely the Airy minima,
can be consistently described by a nucleus-nucleus potential
with a deep real part and a weakly absorptive imaginary part.
The difference in absorption in the two systems is explained by an increased imaginary (mostly surface)
part of the potential in the $^{18}$O$+^{12}$C system.
The relation between the obtained potentials and those reported for the symmetrical
$^{16}$O$+^{16}$O and $^{12}$C$+^{12}$C systems is drawn.

 \vspace{7mm}


\end{abstract}

\begin{multicols}{2}
\section{Introduction}
\label{intro}

Significant progress has been made in the last decade in the 
understanding and determination of  
the nuclear optical potential for light heavy-ion systems. 
In the description of heavy-ion collisions and
the accompanying compound nucleus formation,
the strongly absorptive interaction is usually present.
In certain light heavy-ion collisions,
involving closed or semiclosed shell nuclei,
 the number of open reaction channels is small and the absorption is weaker 
\cite{ha81}. 
Consequently,  resonant and refractive phenomena have been 
observed in these systems.
Recent reviews of
 theoretical and experimental results on
resonant and refractive phenomena can be found in
\cite{reson,betsrab,br97}.
Reported systematic measurements
of the elastic scattering which cover broad angular and
energy ranges and their
optical model analysis have resulted in a
deeper insight into the dynamics of light heavy-ion scattering.
Such progress is closely connected with the observation of refractive effects.
It has been shown that the presence of the nuclear rainbow and,
the associated Airy structure
in elastic scattering reduces
the ambiguities of the optical potential character (shallow or deep).
The existence of systematic studies,
where the change of potential parameters can be
followed as a function of energy, is essential
for the determination of a unique
nucleus-nucleus potential at different energies and in different systems.

The main features of the optical rainbow lie in the refraction and
reflection.
In the case of interference due to the contributions around the 
maximum deflection angle,
higher-order maxima, i.e., Airy structure, appear inside the lighted region.
The observation of the rainbow pattern in heavy-ion collisions is closely
connected with the degree of transparency.
Water is transparent
to visible light,
while the nuclear rainbow is damped by the presence of absorption.
The rainbow scattering appears when the nuclear potential is strong enough
to deflect particles into negative angles (so-called farside).
The combination of these two features, deep real potential and
incomplete absorption, makes possible the observation of distinct
refractive effects like the Airy minima, the rainbow angle and,
the rainbow dark side.

The energy range between 5 and 10 MeV per nucleon
can be defined as an intermediate region lying between that where the
molecular-resonant states show up and
that with the prominent appearance of  the nuclear rainbow.
The experimental results on binary channels reported
up to very recently in the
literature for this energy range are rather scarce.
The refractive studies have mainly involved symmetrical systems,
mostly $^{12}$C$+^{12}$C \cite{stoksad,brc12,moc12} and
$^{16}$O$+^{16}$O \cite{khoa,oertzen,kondoo16}.
To cover the
lack of systematic measurements,
several experiments have been performed at the Strasbourg
Vivitron accelerator, 
which is well suited for studies in this intermediate energy range.
Detailed and complete elastic angular distributions
have been measured for three systems: $^{16}$O$+^{16}$O,
$^{16}$O$+^{12}$C, and $^{18}$O$+^{12}$C.
In this work we concentrate on the study of the nonidentical
systems where the angles beyond 90$^{\circ}$ can be explored directly.

In the literature,
the data for the $^{16}$O$+^{12}$C system are not as complete
and numerous as for identical boson systems.
Relative to the $^{16}$O beam energies of the present work, the
elastic scattering angular distributions have been reported
and analyzed at higher energies: $E_{\rm lab}=$
132 \cite{ogloblin}, 139,
216, and 311 MeV \cite{br21}, 608 \cite{br26}, and 1503 MeV \cite{br175},
and at lower energies: $E_{\rm lab}=$ 24, 35, 45 \cite{ogloblinnew9},
65, and 80 \cite{ogloblinnew10,kondolow} MeV.
The data reported in Refs. \cite{br21,ogloblinnew9,ogloblinnew10} 
cover only limited angular ranges and do not reveal any significant 
refractive features. 
The data at 608 and 1503 MeV have been shown to contain some 
refractive effects displaying sensitivity 
to the features of the real potential. 
Recently, 
the elastic scattering measurements over a wide angular range 
have been published
at energies just above the energy range covered in our study,
i.e. at
$E_{\rm lab}$=132 (re-measured and extended to large angles), 
170, 200, and 230 MeV \cite{ogloblinnew}. 
No refractive study of the $^{18}$O$+^{12}$C system has been 
reported in the literature so far.

Our elastic scattering data and the optical model analysis for 
the $^{16}$O$+^{16}$O  
symmetric system were presented in \cite{mpinpc,mpbormio,mpo16},
and for  the $^{16}$O$+^{12}$C system in \cite{mpoc}.
The preliminary results on the  $^{18}$O$+^{12}$C
scattering we presented in \cite{szrab,szbormio}.
In this article, we report on an additional optical potential analysis
of the $^{16}$O$+^{12}$C and $^{18}$O$+^{12}$C systems.
In particular, we present the nearside-farside decomposition of the deduced
potentials  which allowed one to identify
Airy minima and to follow their evolution as a function of beam energy.
To have a better connection with the higher-energy domain,
in the analysis of the $^{16}$O$+^{12}$C system
we have included the lowest energy (132 MeV) elastic scattering
angular distribution reported in \cite{ogloblinnew}.
The change of the potential due to the addition of  two
extra neutrons in the $^{18}$O$+^{12}$C system
will be explicitly discussed.

\section{Experiments}


The three reactions $^{16}$O$+^{12}$C, $^{18}$O$+^{12}$C, and
$^{16}$O$+^{16}$O were studied under the same 
experimental condition, which will  be briefly described in this section.
More details can be found in Refs. \cite{c12o18,szth}.

The elastic scattering angular distributions were measured  
at laboratory
angles between 5$^{\circ}$ and 20$^{\circ}$ in
steps of 0.5 degree using
the Q3D magnetic spectrometer and its associated detection system.
At larger angles all binary channels were recorded simultaneously using a
fixed kinematical-coincidence setup composed of two
position-sensitive (PS) silicon detectors.
The setup allowed complete mass and $Q$-value
identification of the binary fragments.
PS detectors were placed
on both sides of the beam (15$^{\circ}\le \theta\le 50^{\circ}$ and
$-35^{\circ}\le\theta\le-70^{\circ}$)
at 7.8 cm from the target, a self-supporting carbon film
$\sim20\mu$g/cm$^{2}$ thick.
The $^{16}$O$+^{12}$C and $^{18}$O$+^{12}$C elastic scatterings
were measured at
beam energies
$E_{\rm lab }(^{16}{\rm O})$=62, 75, 80, 94.8, 100, 115.9, and 124 MeV
and, at
$E_{\rm lab}(^{18}{\rm O})$=66.2, 85, 100, and 120 MeV, respectively.
The electronics and data acquisition system of the Q3D spectrometer
with the proportional counter in its focal plane and
of the PS detectors
in the kinematical-coincidence mode were independent of each other.
The two detection systems
had an overlapping angular region.
The obtained center-of-mass elastic scattering
angular distributions span the angular range
between 10$^{\circ}$ and more than 140$^{\circ}$ for both
systems studied with good statistics at all measured energies.

All angular distributions and the optical model description of the
$^{16}$O$+^{12}$C system  were reported in \cite{mpoc}.
In the present work we intend  to study in more detail  the
connection with the higher-energy region.
Thus, we are not going to discuss the results obtained at
the two lowest energies,
62 and 75 MeV.
In fact,
it has recently been shown that at even lower energies
the deep real potential describes both the observed structure in the
fusion excitation function and the gross features of elastic angular distributions
\cite{kondolow}.

Resonant phenomena in heavy-ion reactions are usually restricted to  systems
composed of the so-called $\alpha$-particle nuclei.
Considering the presence of resonances as a signature of weak absorption,
one would expect the refractive effects to appear in
the higher-energy domain for the same systems.
A description of all details that appear in
angular distributions where the resonant effects are superimposed
on the gross structure is unlikely to be achieved with a mean field approach,
i.e., the optical model.
This is the case
with the $^{18}$O$+^{12}$C system at $E_{\rm lab}$=66.12 MeV,
the lowest energy of the present measurement.
At this energy a broad resonance has been reported and the
energy spectra and angular distributions of the binary channels
were discussed in  \cite{c12o18,c14o16}. 
 
\section{Optical model potential}

In the analysis of the $^{16}$O$+^{16}$O 
\cite{mpinpc,mpbormio,mpo16} and $^{16}$O$+^{12}$C \cite{mpoc} 
systems 
several choices were tried for the real part of the potential. 
Our understanding of the properties and microscopic grounds of the optical
potential is 
largely based upon the folding model using a realistic 
nucleon-nucleon effective interaction \cite{folding}. 
The energy- and density-dependent  interaction was used to
generate a microscopic potential, which
together with a phenomenological imaginary term, successfully describes
our $^{16}$O+$^{16}$O and $^{16}$O$+^{12}$C data as well
as the collisions in slightly higher
and lower energy domains \cite{oertzen,kondolow}.
The resulting folded potentials are strongly attractive,
and weakly energy dependent.
It has been shown (see Fig. 1 in \cite{mpoc})
that the square of the Woods-Saxon
form factor (WS2) with a suitable choice of radius and diffuseness
gives a shape that is very close
to the folded potential.
Indeed, equivalent fits to the data are
found with either the folded or the WS2 potentials.
In the present case, the real part of the potential is taken in the WS2
form:
\begin{eqnarray}
V(r)&=&-V[f_{V}(r)]^{2},
\label{eqre}  \\  \nonumber \\
f_{V}&=&[1+exp(\frac{r-R_{V}}{a_{V}})]^{-1}.  \nonumber
\end{eqnarray}

\noindent
The shape of the folded potential  for the $^{16}$O$+^{12}$C system
is closely reproduced by Eq. (\ref{eqre}) with
the radius $R_{\rm V}=$4.0 fm and the surface diffuseness $a_{\rm V}=$ 1.4 fm.
These values were kept constant throughout the best-fit
automatic parameter searching procedure.
In the energy range considered,
no significant improvement of the fit was obtained by allowing
$R_{\rm V}$ and $a_{\rm V}$ to vary freely.
Slightly different values of the radius and diffuseness
of the real potential
for $^{18}$O$+^{12}$C were adopted
owing to the
addition of the two neutrons.
They were kept constant throughout the
fitting procedures at the values $R_{\rm V}=$4.08 fm and  $a_{\rm V}=$ 1.38 fm.

Two choices were used for the imaginary potential, a pure "volume" term
of the Woods-Saxon (WS) type ($W_{\rm WS}$) and
a volume term together with an additional "surface" term
($W_{\rm WS2}+W_{\rm D}$).
For the second choice of the imaginary potential
we adopted the WS2 form factor
for the volume term, and
 the radial derivate of the WS form
for the additional surface term.
The imaginary potentials are thus given by the following equations:
\begin{eqnarray}
W(r) & =&W_{\rm WS}(r)=-iW[f_{W}(r)], \label{eqws1} \\  \nonumber\\
W(r)&=&W_{\rm WS2}(r)+W_{\rm D}(r)\nonumber\\
&=&-iW[f_{W}(r)]^{2}-iW_{D}f_{D}(r), \label{eqws2} \\ \nonumber \\
f_{W}&=&[1+exp(\frac{r-R_{W}}{a_{W}})]^{-1}, \nonumber \\
f_{D}(r)&=&-4a_{D}\frac{d}{dr}[1+exp(\frac{r-R_{D}}{a_{D}})]^{-1}. \nonumber
\end{eqnarray}

\noindent
In the following text, the optical potentials defined by  
Eqs. (\ref{eqre}) and (\ref{eqws1}) will be called parameterization
$P(W_{\rm WS})$ and
those defined by Eqs. (\ref{eqre}) and (\ref{eqws2})
will be called parameterization $P(W_{\rm WS2}+W_{\rm D})$.
The fits were obtained through the automatic search option in the 
program PTOLEMY \cite{ptol}.

The usual ${\mathcal{X}}^{2}$ criterion was used to judge the quality of agreement with the data:
\begin{eqnarray}
{\mathcal{X}}^{2}=\frac{1}{N_{\sigma}-N_{\rm P}} \sum^{N_{\sigma}}_{i=1}
 \frac{(\sigma_{\rm th}^{\rm i}-\sigma_{\rm ex}^{\rm i})^{2}}{(\Delta \sigma_{\rm ex}^{\rm i})^{2}}.
 \\ \nonumber
\end{eqnarray}
The $\sigma_{\rm th}$ and $\sigma_{\rm ex}$ are theoretical and 
experimental cross sections, $\Delta \sigma_{\rm ex}$ are the uncertainties in the 
experimental cross sections, $N_{\sigma}$ is the total number of 
angles at which measurements were carried out,
and $N_{\rm P}$ is the number of free fitting parameters.
Two choices for the cross section uncertainties were considered.
We either used the experimental cross section uncertainties or
a fixed percentage of the experimental
cross section for all measured angles.
Compared with forward angles,
the experimental cross-section uncertainties can be relatively large at
intermediate and backward angles owing to the smaller cross section.
Thus, the searching procedure does not give sufficient weight to those
angles at which the refractive effects are expected to be observed.
We have found that using a uniform percentage (10$\%$)
a better fit at intermediate and backward angles can be obtained.

For an interaction potential $U_{\rm E}(r)$ between
nuclei with nucleon numbers $A_{1}$ and $A_{2}$, the volume integral

\begin{eqnarray}
J_{\rm U}(E)=-\frac{4 \pi}{A_{1}A_{2}}\int U_{\rm E}(r)r^{2}dr
\label{integral}
\end{eqnarray}
\noindent
is a sensitive measure of the potential strength.
This definition
 applies to the real and imaginary parts of $U_{\rm E}(r)$.
Generally, the optical model analyses can give
several discrete families of real potentials that fit
an angular distribution equally well.
It has been shown that the volume integrals are valuable tools for
 classifying different optical model families, especially
in terms of global systematics, when different systems are compared.

The simple optical model potential is a local,
angular-momentum independent potential.
Thus, effects like the elastic
transfer, an $\alpha$-transfer in  $^{16}$O$+^{12}$C or
a  $^{6}$He-transfer in $^{18}$O$+^{12}$C,
which can introduce additional structure in the angular distribution,
are not included.

The obtained angular distributions are complicated to interpret
owing to the contributions and interferences of many involved partial waves.
For this purpose,
the nearside-farside decomposition technique, first presented
by Fuller \cite{fuller}, is a helpful method for
interpreting the optical model results.
The nearside-farside decomposition separates the trajectories
which originate from two different sides of the scattering potential.
By convention, the nearside and farside correspond to
classical trajectories
with positive and negative deflection angles, respectively.
The interference between nearside and farside trajectories
will lead to the Fraunhofer diffraction pattern,
and the interference between two farside components
(the one more peripheral and the other more in the interior)
will lead to the rainbow pattern.
The refractive effects, namely the Airy minima and maxima,
will therefore be present in the farside component.
The crucial condition for the observation of
refractive effects is incomplete absorption.
The absorption should be weak enough ("transparency")
to allow the inner
farside component to be effective.
The appearance of Airy oscillations in an angular distribution is a
direct indication of  interferences between waves which are bent by the nuclear
mean field and provides a unique information on
the potential at small interaction radii.

\section{Elastic angular distributions}
\subsection{Optical model description}

Figures \ref{elo16} and \ref{elo18} show the data and the 
phenomenological model fits (thick-solid curves)
of the angular distributions
for the $^{16}$O$+^{12}$C and
$^{18}$O$+^{12}$C systems, respectively.
The parameters obtained using the fitting procedure are listed in
Table \ref{tws2} (parameters $P(W_{\rm WS2}+W_{\rm D})$ which
are used on the left panels of both figures) and
Table \ref{tws1} ($P(W_{\rm WS})$ and
the right panels of Figs. \ref{elo16} and \ref{elo18}).

The main
features of the measured angular distributions, namely,
 the forward fine Fraunhofer diffractive oscillations and the
broad structure at larger angles
 are well described by the calculation.
An increase in cross sections at large angles is predicted by the
calculations, but is smaller in some cases than
observed experimentally.

In our previous analyses of the $^{16}$O$+^{16}$O system
\cite{mpo16}, it was shown that for the data at the three highest
energies (124, 115.9, and 103.1 MeV) an acceptable
description was possible using just a volume term for
the imaginary part.
The  fits  at lower energies
(lower than 100 MeV)  required
inclusion of a surface
imaginary term.
To be consistent, in the present analyses, the $^{16}$O$+^{16}$O
 data at all  energies were analyzed using both imaginary terms,
volume and surface.
The resulting imaginary potential is mostly of the volume type at higher energies.
At lower energies the imaginary potential has a sharp-edged volume term and a surface term which
peaks outside the volume one (see the top-right panel of Fig. \ref{pot-all}).
The initial set of parameters for the fits of the
$^{16}$O$+^{12}$C system was the one which gives the same volume
integrals of the real and imaginary parts (Eq. \ref{integral})
as those obtained in the fitting procedure for
the $^{16}$O$+^{16}$O system.
The final difference between the volume integrals of the
$^{16}$O$+^{16}$O and $^{16}$O$+^{12}$C
systems at 124 MeV is $\sim$10$\%$.
In the optical model analysis of the $^{16}$O$+^{12}$C system
presented in \cite{mpoc}, at all measured energies
the folding and WS2 terms were used for the real potential and
the sum of the volume (WS) and surface terms for the imaginary
potential.
It has to be noted that the imaginary volume term in \cite{mpoc}
was of the WS type, while in the present analysis it is of the WS2 type when the
surface term is included,
which results in slightly different parameters.
The main goal of this work is to establish a connection
with the higher-energy
domain and to identify the Airy minima and their order
using the
nearside-farside decomposition technique.
To fulfill this task, we included in our study
the angular distribution at 132 MeV \cite{ogloblinnew}.
The optical model analysis presented in \cite{ogloblinnew} was 
performed
by using for the real part either the folding or the phenomenological Woods-Saxon potentials. 
For the imaginary part, only a volume term was used
(WS at all energies and WS2 at several energies).
Such a potential, with only a volume imaginary term,  describes the main features
of the measured distributions, but fails to describe the 
backward-angle oscillations which become more important as energy decreases.
Such a potential underestimates the 
observed increase of the yield  
at backward angles.

Actually, a satisfactory description of the whole structure observed 
 in the   $^{16}$O$+^{12}$C system in the energy range considered, 
required the inclusion of an 
imaginary surface term. 
To justify this statement, the analysis with and without the
imaginary surface term is consistently carried out and presented. 
Indeed, 
the distributions calculated without the surface 
term resulted in too little structure at large angles 
(Figs. \ref{elo16} and \ref{elo18}, right panels).
In addition, 
the predicted increase of yield at backward angles (in agreement with data) is 
less pronounced when the cross section is calculated using the 
imaginary volume term only. 
It seems that for a certain energy range, the data cannot be 
sufficiently well described without the surface term  and that the 
$^{16}$O$+^{12}$C and $^{18}$O$+^{12}$C systems 
at the energies considered here are within  this range.

For the $^{18}$O$+^{12}$C system,
the fits were obtained starting from the parameters of the 
$^{16}$O$+^{12}$C system. 
The results of this procedure at an energy of  100 MeV,
can be seen in Fig. \ref{elo16o18}.
The top panel shows the best fit of the
$^{16}$O+$^{12}$C system with two different sets of parameters,
using the sum of the volume and surface terms
(solid curve) and using the volume term only (dashed curve).
These sets of optical potential parameters were
 used to calculate the angular
dependence for the $^{18}$O+$^{12}$C system.
 The results of the calculation are
superimposed on the $^{18}$O+$^{12}$C data in the middle panel
of Fig.  \ref{elo16o18}.
Agreement between
measured and calculated values is reasonable  at forward angles,
but the calculated angular distributions cannot describe
the two orders of magnitude smaller yield in
the $^{18}$O+$^{12}$C  system at larger angles.
It is worth noting that even without fitting, the number of
oscillations in the calculation obtained using the
imaginary term composed of the volume and surface terms (solid curve)
agrees with the measurement throughout the  angular range and
over eight orders of magnitude of the differential cross section.
The bottom panel of Fig. \ref{elo16o18} shows the results of
the fitting procedure.
The differences between the potentials of the two systems
are presented in Fig. \ref{poto16o18e100}:
The left-bottom panel refers to the $P(W_{\rm WS})$ parametrization
and the right-bottom panel to the $P(W_{\rm WS2}+W_{\rm D})$ one.
The real part remains almost unchanged, but the imaginary volume integral
increases substantially from $^{16}$O$+^{12}$C to $^{18}$O$+^{12}$C.
To describe the decrease of the yield at backward angles,
the  absorption has to be increased.
In the imaginary part of the potential which is composed of the volume and surface terms,
the surface term now becomes dominant at large interaction radii.

The obtained potentials have deep real parts.
When the additional surface term in
the imaginary potential is included, the volume term
has a smaller radius than the surface term and  as a consequence,
the volume imaginary potential has an almost square profile whose
detailed shape depends on the system and energy.

The volume diffuseness of the imaginary potential tends to be very
small when the surface imaginary term is included in the potential.
The volume imaginary diffuseness
could be fixed at some small values, such as 0.1 fm,
without any significant change of the fits,
but we rather report the values resulting from the automatic search.

The importance of the volume integrals can be shown
by comparing the
results of the two different parameterizations,  $P(W_{\rm WS})$
and  $P(W_{\rm WS2}+W_{\rm D})$.
The potentials which belong to $P(W_{\rm WS})$
and  $P(W_{\rm WS2}+W_{\rm D})$
 at 100 MeV are shown in Fig. \ref{poto16o18e100}: left-top panel for
the $^{16}$O$+^{12}$C and right-top panel for the $^{18}$O$+^{12}$C systems, respectively.
The dashed curves correspond to   $P(W_{\rm WS})$,
while the solid curves correspond to the  $P(W_{\rm WS2}+W_{\rm D})$.
The two solutions give the same
(within a few percent) volume integrals (see Table \ref{tintegrali}).
In the imaginary part, when the individual surface and volume terms
are considered, both terms are important.
As mentioned above, the common characteristic of the
$W_{\rm WS2}+W_{\rm D}$  solutions is that
the volume imaginary term tends to have a sharp edge and a smaller
radius than the surface term.
The precise balance between the two imaginary terms  is not so well
established as the values of the integrals.
The imaginary potential
consisting of the volume term only,
tends to  smooth out the potential pocket of the  $P(W_{\rm WS2}+W_{\rm D})$
in a way as to keep the imaginary volume integral about the same.

The volume integrals of the real potentials are very similar for both
parameterizations
and also for the phenomenological and the folding potentials.
In \cite{ogloblinnew} two families of discrete WS sets were obtained
using the fitting procedure.
The obtained real WS potentials have different depths and similar radii and consequently
different volume integrals.
At  lower energies  both WS sets describe the data equally well,
but the deeper potentials (the WS2 set in \cite{ogloblinnew})
fail to reproduce the right
order of Airy minima at higher energies.
This is clearly visible in the data at 200 MeV of
\cite{ogloblinnew} where the first Airy minimum and
the rainbow angle are observed.
Comparing the values of the volume integrals and the numerical values of the
potentials, we may conclude that
the real potential obtained in this work belongs to the same family
(the WS1 set in \cite{ogloblinnew}).
The selection of a unique potential family results in
identification of the order of Airy minima. Furthermore, it makes 
possible to follow the evolution of the Airy structure with energy.

\subsection{Nearside-farside decomposition}

The nearside-farside decomposition is particularly helpful in the  
intermediate energy range of our study. 
In Figs. \ref{elo16} and \ref{elo18}, the thin-solid and dashed curves
show the nearside-farside 
decompositions of the obtained optical model fits of 
the $^{16}$O$+^{12}$C and the 
$^{18}$O$+^{12}$C systems, respectively.

The angular distributions at the highest and the lowest  energy
of the $^{16}$O$+^{12}$C system 
(top and bottom panels of Fig. \ref{elo16}) show different patterns. 
The fit and calculation for the 132 MeV angular distribution 
display the features which are common to this and similar systems at higher energy.
We can describe the angular distribution by Fraunhofer oscillations at forward angles, 
glory effects at backward angles,  and intermediate structures, namely the
deep minimum at 80$^{\circ}$, as an Airy oscillation. 
The nearside-farside crossover takes place around 25$^{\circ}$.
 Beyond this angle, the angular distribution is farside dominated and the structure 
of the distribution is  
refractive in its origin, i.e., it is the result of the interference between 
the two farside subamplitudes.
By definition, the nearside and farside amplitudes are equal at 180$^{\circ}$
and their interference produces strong and rapid 
oscillations (glory effect) in the angular distributions as we approach the largest angles. 
The situation changes at lower energies. 
The nearside-farside crossover moves to larger angles and 
 the Fraunhofer oscillations also extend up to large angles, 
they are superimposed on broad oscillations of refractive origin. 

The comparison between the decompositions of the two different 
sets of parameters of the imaginary potential
(left and right panels of Fig. \ref{elo16})
shows that the obtained
farside components are very similar, while
the nearside components exhibit quite different behavior.
As already stressed, the  
refractive effects appear in the farside components and, 
omitting the different oscillatory pattern at very backward angles,  
the location of the Airy minima of the two different parameterizations  
remains the same. 
Such behavior clearly shows that the Airy structure is determined
entirely by the real potential.

The nearside and farside components are of comparable 
strengths at intermediate angles 
for the  $P(W_{\rm WS2}+W_{\rm D})$.
The nearside  recovers its strength
beyond the nearside-farside crossover.
For the decomposition of the calculations obtained with  the $P(W_{\rm WS})$
 and shown on the right panels,
the farside components behave more as was observed at higher
energies, i.e., the angular distribution is farside dominated,
whereas the nearside is more than two orders of magnitude smaller.

The angular distribution measured at 115.9 MeV is
a typical example of the cross section
in this intermediate energy range.
The nearside-farside crossover takes place at 50$^{\circ}$, 60$^{\circ}$,
and 100$^{\circ}$ for the  $P(W_{\rm WS2}+W_{\rm D})$.
The nearside and farside amplitudes become interlaced owing to their similar  strengths and
intersect at several points.
The nearside-farside crossover takes place at 27$^{\circ}$ for the 
 $P(W_{\rm WS})$ and above this angle the distribution 
is farside dominated. 
Again at backward angles, the interference between the nearside 
(coming partly from the farside component after passing through 180$^{\circ}$ 
and going around) and farside components
is clearly visible in the angular distribution. 
Looking only at the farside amplitude, two 
Airy minima appear in the amplitude ($A_{2}$ near 95$^{\circ}$  
and $A_{3}$ near 60$^{\circ}$) for both parameterizations. 
The data themselves show a wide minimumlike structure at these angles.

Owing to the stronger absorption in the $^{18}$O$+^{12}$C system
(see Fig. \ref{elo18}), 
even at our highest energy (120 MeV), the nearside and
farside amplitudes are of similar strength throughout the
whole measured angular range (top-left panel in Fig. \ref{elo18})
for  the $P(W_{\rm WS2}+W_{\rm D})$, whereas for the
$P(W_{\rm WS})$ the nearside-farside crossover is shifted
toward larger angles (top-right panel in Fig. \ref{elo18}).
The similarity of the nearside and farside strengths prolongates the
appearance of the diffractivelike oscillation up to large angles.
Moreover, the nearside-farside crossover moves even to larger angles as energy decreases.
In the description of these  data we have to keep
in mind that the farside which turns over at 180$^{\circ}$ becomes the nearside,
and that the structure observed at backward angles can be refractive in its origin \cite{mpo16,mpoc}.
In spite of this angle-extended Fraunhofer diffraction pattern
the Airy minima can be discerned
by considering only the farside amplitudes.
In the raw data themselves the refractive effects are observed rather
as irregularities in the oscillation pattern.
 Owing to the stronger absorption in the $^{18}$O$+^{12}$C system, the
nucleus-nucleus interaction transparency is
reduced and the interference between the two farside
subamplitudes does not dominate the angular distributions
in the energy range studied.
The nearside-farside decomposition with the $P(W_{\rm WS2}+W_{\rm D})$
shows that the experimental distributions are
nearside dominated, i.e., the diffraction is
important throughout the measured angular range.
We have to keep in mind that the $P(W_{\rm WS2}+W_{\rm D})$
better describes the observed structure at larger angles.
Nevertheless, even if the deep minima of refractive origin
in the angular distribution are not observed,
the description of the structure is within the refractive picture,
i.e., the deep real potential is required.
We want to emphasize that whenever a choice among alternative solutions
of the optical potential had to be made, the main criterion was
the regularity (monotonic variation) of the parameters.
The regularity was based on the expectation that the potential
should not change rapidly with energy and for neighboring systems.
The resulting potential parameters which show such  a regularity
allow the connection of the observed structure in the elastic
angular distributions of the $^{18}$O$+^{12}$C system with the refractive effects,
as observed in the neighboring and more
"transparent" $^{16}$O$+^{16}$O and $^{16}$O$+^{12}$C systems.
Since the refractive effects are not so prominent in
the $^{18}$O$+^{12}$C angular distributions, the
 optical model ambiguities remain to a certain extent.

We may conclude that in the intermediate energy range,
between 5 and 10 MeV per nucleon, refractive effects
are an important ingredient of the global structure of an angular distribution.
The ambiguities in the determination of these refractive effects are closely
related to the  degree of absorption in the system.
Even if the refractive effects are masked,
the structure observed in  the measured angular distributions
can be at least partially  explained as being of refractive origin.
The stronger the absorption, the more masked are the refractive effects  
and their description becomes more complicated and less accurate.
Alternative explanations of the observed structures in angular
distributions at this intermediate energy range can be helpful.

Another possibility is the decomposition into the
barrier- and internal-wave (B/I) components \cite{ohkubo1,ohkubo2}. 
The B/I decomposition makes sense if the real part of the potential is 
deep enough for the effective potentials to display a "potential pocket"  
and if the absorption is incomplete. 
Recently, the B/I decomposition has been applied to our data,
at all energies for  
$^{16}$O$+^{16}$O \cite{ohkubo1,ohkubo2}, and also at one representative energy 
for our nonidentical systems $^{16}$O$+^{12}$C and $^{18}$O$+^{12}$C \cite{ohkubo2}.
Within the B/I decomposition, the Airy minima are explained
as an interference of the barrier-wave and internal-wave subamplitudes.
The results of the B/I decompositions for the lower and higher
energy range of our $^{16}$O$+^{16}$O data differ only in
the relative importance of the two components.
Inclusion of the surface imaginary term
modifies  the internal contribution only slightly,
but produces the increase of the barrier cross section at large angles,
which results in the increase of the yield and the presence
of  oscillations in the cross section.
The prominence of the Airy minimum in the 132 MeV angular distribution of
$^{16}$O$+^{12}$C is due to similar magnitudes of the
internal-wave and barrier-wave
components  in the angular range around 80$^{\circ}$.
Owing to the stronger absorption in the $^{18}$O$+^{12}$C system,
the internal-wave cross section is about three orders of magnitude
lower than in the $^{16}$O$+^{12}$C system,
which almost completely smears out
the appearance of Airy minima in the experimental
angular distributions.

Very recently a semiclassical analysis of the Airy-like pattern has been presented \cite{anni}.
The oscillations in the farside subamplitude are explained as
a result of interference of the first and second term of a
multireflection expansion of the scattering function.
The physical contents  and the obtained decomposition are very similar to
the results of the B/I decomposition.

\subsection{Interpretation of the results in terms of Airy minima}
   
The rainbow and associated Airy oscillations in the scattering 
appear in the farside component. 
The Airy structure is controlled by the real potential and its 
appearance is often obscured by the presence of absorption. 
As the energy increases, the rainbow and Airy minima move forward in angle. 
The identification of the structures of refractive origin can be simplified by
using a reduced imaginary potential. 
Figures \ref{nfo16} and  \ref{nfo18} show the farside amplitudes 
 with a $50\%$ reduced imaginary strength  
of the $^{16}$O$+^{12}$C and $^{18}$O$+^{12}$C systems, respectively. 
The real potential is weakly dependent on energy 
and the smooth energy evolution of the Airy minima can be 
observed in the calculated distributions in Figs. \ref{nfo16} and \ref{nfo18}.

Let us first discuss the results for the  
$^{16}$O$+^{12}$C system (Fig. \ref{nfo16}).
In Refs. \cite{ogloblinnew,sattobepub} the strong minimum at 
80$^{\circ}$ in the measured 132 MeV angular distribution  is identified
as the second Airy minimum $A_{2}$. 
The minimum at 55$^{\circ}$ 
was identified as $A_{3}$ and that at $35^{\circ}$ as $A_{4}$.
As already discussed, the real potentials obtained
in our analysis for the $^{16}$O$+^{12}$C system
belong to the same family as the real potentials
reported in \cite{ogloblinnew}.
Following the systematics of the order of Airy minima,
the minima in the angular distribution at 124 MeV
appearing at angles
88$^{\circ}$, 60$^{\circ}$, and 40$^{\circ}$
are identified as $A_{2}$, $A_{3}$ and $A_{4}$.
At 100 MeV the $A_{2}$ minimum moves to 120$^{\circ}$,
$A_{3}$ to 80$^{\circ}$, and $A_{4}$ to 50$^{\circ}$.
At very forward angles, around 30$^{\circ}$, the fifth
Airy minimum can also be discerned in the farside component at 100 MeV.
Of course,  it is not easy to see the remnants of all
these minima in the actual data, especially for the higher-order minima
appearing at more forward angles where strong
Fraunhofer oscillations are dominant.

No detailed elastic angular distributions
at the $^{18}$O$+^{12}$C system
for energies higher than those in the present study (where one expects a more
favorable situation for the identification of the Airy minima and their order)
have been reported so far.
In the identification of the order of Airy minima,
we assume a similar behavior in the $^{18}$O$+^{12}$C and  $^{16}$O$+^{12}$C systems.
Therefore, it is
likely that the $A_{2}$ and $A_{3}$ minima appear in
$^{18}$O$+^{12}$C at energies around 100 MeV as well.
According to this assumption, the minimum around 105$^{\circ}$ in  the
120 MeV angular distribution is labeled as $A_{2}$, and
the minima at 75$^{\circ}$, 50$^{\circ}$, and at the very forward angle
of $30^{\circ}$ as $A_{3}$, $A_{4}$, and $A_{5}$ (see Fig. \ref{nfo18}).
The minima move to larger angles at lower energies and
at 100 MeV the $A_{3}$ moves to
105$^{\circ}$, $A_{4}$ to 72$^{\circ}$, and $A_{5}$ to 45$^{\circ}$.
At 85 MeV, even the $A_{6}$ can be discerned at a very forward angle.

The positions of the Airy minima as a function of center-of-mass energy
for both systems studied are shown in Figs. \ref{airyo16} and \ref{airyo18}.
A very regular parabolic behavior is observed.

The angles at which the minima take place in the
farside amplitudes do not
depend on the imaginary potential strength and no
important angular shift is observed if the absorption is switched on.
The depth of the minima in the farside
changes with absorption.
The absorption dictates which minimum will be dominant in the
farside amplitude.
The stronger the absorption, the deeper  the minima will be at larger angles
and conversely at smaller angles.
It has to be noted that
the interference between the volume and surface components
introduces some additional small oscillations around the Airy minima. 
This speaks in favor of the conclusion that the volume and surface absorptive components
play different roles.

\subsection{Potentials, volume integrals, and global systematics}
In this section we present all optical potentials obtained for the 
four systems considered in our study: 
$^{16}$O$+^{16}$O \cite{mpinpc,mpbormio,mpo16},
$^{16}$O$+^{12}$C \cite{mpoc,szrab,szbormio},  
$^{18}$O$+^{12}$C \cite{szrab,szbormio},  and $^{12}$C$+^{12}$C
\cite{moc12,szbormio,szjpg}.
 Figure \ref{pot-all} shows the 
energy and system variations of the obtained potentials (real
and imaginary parts) at representative  energies of the energy range studied.

In all systems, a deep real part is required for a good
description of the elastic angular distributions. 
The observed structure in the distributions is explained 
through refractive effects. 
The imaginary part changes from system to system. 
While the inclusion of the
surface term was  needed only for the lower energy range in $^{16}$O$+^{16}$O,
the $^{16}$O$+^{12}$C system 
required a surface term for all
energies, and without it the structure at larger angles cannot be 
well described.
The surface term becomes even more important in the 
obtained imaginary potential of the 
$^{18}$O$+^{12}$C system.  
When the surface term is needed, the absorptive
potential has a characteristic shape: a volume term which tends to 
have a small diffuseness  plus a Gaussian-like surface peak. 

The most typical examples of such a shape are the optical
potentials for $^{16}$O$+^{16}$O at 75 MeV, for $^{16}$O$+^{12}$C
at 124 MeV, and for  $^{18}$O$+^{12}$C at all energies studied.
Within a simple interpretation, we can associate the
volume term with the absorption due to fusion and related processes,
and the surface term with more direct reactions.
These different shapes of the
absorptive potential can be explained as a consequence of the  
evolution of the 
reaction mechanisms which change from the fusion type to a more 
direct type of reaction. 
This enhancement of the surface integral is a signature of a
larger number of open direct reaction channels and is in agreement with the number
of open channel calculations \cite{ha81}.  
Complementary studies of the distribution of the incident flux into the different 
available exit binary channels, as well as of the underlying 
reaction mechanisms are actually underway for the
$^{16}$O$+^{12}$C and $^{18}$O$+^{12}$C reactions.

Table \ref{tintegrali} lists the real and imaginary volume integrals of the potentials obtained
for the oxygen+carbon systems studied and for the two adopted sets of parameters. 
The real and imaginary volume integrals display a smooth behavior with energy.
The real volume integral generally decreases with energy, while the
imaginary volume integral increases. 
Such behavior is in agreement with the dispersion relation predictions. 
Although the shapes of the imaginary potential are radically 
different for the two sets of parameters, the full imaginary volume integrals 
(columns 5 and 7 of Table \ref{tintegrali})  are about the same.

In the review article of Brandan and Satchler \cite{br97}, 
the values of the real and imaginary volume integrals as a function of projectile energy 
per nucleon for the potentials that fit the 
$^{12}$C$+^{12}$C, $^{16}$O$+^{16}$O, and $^{16}$O$+^{12}$C data have been extracted
(see Fig. 6.7 of \cite{br97}). 
It is shown that the different real potentials all have  similar volume integrals and that
the imaginary volume integrals show a smooth behavior as a function of energy 
in spite of various parameterizations adopted for the imaginary potentials. 
It is quite remarkable to notice that the volume integrals reported in our work closely follow
the global systematics evidenced in \cite{br97}.

\section{Conclusion}

In this paper we have presented the elastic scattering data of the 
$^{16}$O$+^{12}$C and $^{18}$O$+^{12}$C reactions, 
together with the optical model analysis. 
We have used the nearside-farside decomposition technique to 
interpret the complicated features of the angular distributions. 
The elastic angular distributions show not only the usual 
Fraunhofer diffraction pattern, but also, at larger angles, 
refractive effects in the form of nuclear rainbow Airy structures. 
The main features of the measurements are very well 
described by the optical model fits.  
A deep real part is required for the description of refractive effects. 
The imaginary potential is weakly absorptive and reflects 
the presence of incomplete absorption.
The two studied systems,
$^{16}$O$+^{12}$C and $^{18}$O$+^{12}$C,
have different shapes of the imaginary potentials and  different 
values of the volume imaginary integrals. 
The volume imaginary integral increases for the 
$^{18}$O$+^{12}$C system, mostly in the imaginary surface part.
The enhancement of the surface integral is a signature of a 
larger number of open direct reaction channels.
This conclusion is supported by the study of the 
inelastic and transfer channels, as well as by the number of  open channel calculations.

The inclusion of the surface imaginary term provides the 
needed increase of 
the yield at large angles, but also defines a characteristic shape of the imaginary potential. 
When the surface term is required by the data, the imaginary volume term 
tends to have a small difusseness and a smaller radius than the surface imaginary term.
One can imagine that the reflection from a potential with  such a profile may 
produce additional interferences resulting in more structured angular 
distributions at larger angles. 
As has been mentioned, the effects of elastic transfer could also 
cause an increase of the yield at the largest angles and additional 
structure in the angular distributions. 
The specific imaginary potential obtained here can be understood through 
coupling effects which should have an impact on the elastic angular distribution.
An additional analysis like the explicit inclusion of elastic transfer 
could provide a  better understanding of
the origin of the large angle structures, but it is unlikely that this
would appreciably change the results for the real part of the potential.
Therefore, the most important conclusion about the depth of the real potential,
the potential which describes the refractive effects and 
defines the position of the Airy minima, will remain unchanged. 

The appearance of refractive structures permits an unambiguous
determination of the optical potential.
Within this analysis we do not claim that the obtained
potentials are unique, energy by energy.
What we believe as unique is the regularity and systematics of the
potentials as a function of energy
and also of the target and projectile matter distribution.
The obtained potentials for the light heavy-ion systems studied
have a deep real part and a weakly absorbing imaginary part. 
In spite of the differences in the shape of the imaginary potentials for the 
$^{12}$C$+^{12}$C, $^{16}$O$+^{16}$O, $^{16}$O$+^{12}$C,
and $^{18}$O$+^{12}$C systems 
(see Fig. \ref{pot-all}) the volume imaginary integrals agree with the 
global systematics (evidenced in \cite{br97}). 
The systematics is based on the refractive nature of 
light heavy-ion collisions and is in agreement
with the dispersion relation predictions.
The obtained volume integrals, both real and imaginary,
agree with potentials which fit the higher and lower energy data.

We may thus conclude that in the intermediate energy region,
between 5 and 10 MeV per nucleon, the refractive effects,
although masked partially by diffraction, are still strong enough to permit a
simple mean-field optical-model explanation of the main structures of
the elastic scattering cross sections.

\section{\small ACKNOWLEDGMENTS}
\label{ack}

We are indebted to  A.A. Ogloblin and his group for making their experimental angular distributions
available to us.
We would also like to thank Dao T. Khoa for providing us with 
his nearside/farside decomposition codes as well as with a 
UNIX adapted version of the code PTOLEMY.

 \newpage

\begin{table}
\caption{Phenomenological potentials, the real part is a WS2 term and the imaginary part is a sum
of the WS2 (volume) plus WSD (surface) term.
$V$ and the subscript $V$ stand for the real term,  $W$   
and the subscript  $W$   for the volume imaginary term and the subscript $D$ for
the surface imaginary term. 
The $a_{\rm W}$ could be fixed arbitrarily at some small 
value, such as 0.1 fm, without much influence on the fits, 
but we prefer to report the values 
resulting from the automatic search.}
\label{tws2}
\begin{tabular}{ccccccccc}
\multicolumn{9}{c}{
$^{16}$O$+^{12}$C$~~~$ $R_{V}$=4 fm,  $a_{V}$=1.4 fm}\\\hline
$E_{\rm lab}$ & $E_{\rm c.m.}$ & V & W & R$_{W}$ & a$_{W}$ & W$_{D}$ & R$_{D}$
& a$_{D}$ \\ 

[MeV] & [MeV] & [MeV] &  [MeV] & [fm] & [fm] & [MeV] & [fm]
& [fm] \\
\tableline
132   & 56.6 & 292  & 24.6 & 3.190 & 0.420 & 11.3 & 4.880 & 0.640   \\
124   & 53.2 & 296  & 14.7 & 4.399 & 0.170 & 9.3  & 5.985 & 0.453  \\
115.9 & 49.7 & 288  & 16.0 & 4.464 & 0.111 & 7.1  & 6.097 & 0.460  \\
100   & 42.9 & 288  & 10.5 & 5.466 & 0.190 & 3.8  & 6.640 & 0.440  \\
94.8  & 40.6 & 285  & 11.4 & 5.351 & 0.156 & 3.6  & 6.708 & 0.360  \\
80.0  & 34.3 & 278  & 13.9 & 5.256 & 0.170 & 2.5  & 6.849 & 0.438 \\
\tableline
\multicolumn{9}{c}{
$^{18}$O$+^{12}$C$~~~$ $R_{V}$=4.08 fm,  $a_{V}$=1.38 fm}\\
\tableline
$E_{\rm lab}$ & $E_{\rm c.m.}$ & V  & W & R$_{W}$ & a$_{W}$ & W$_{D}$ & R$_{D}$
& a$_{D}$  \\

[MeV] & [MeV] & [MeV] & [MeV] & [fm] & [fm] & [MeV] & [fm]
& [fm] \\
\tableline
120 & 48 & 298 &  22.6 & 3.910 & 0.059 & 11.6 & 5.750 & 0.528 \\
100 & 40 & 310 &  22.0 & 4.100 & 0.047 & 11.0 & 5.730 & 0.540 \\
85  & 34 & 326 &  20.7 & 4.157 & 0.055 & 10.5 & 5.743 & 0.551 \\
\end{tabular} 
\end{table}

\begin{table}
\caption{Phenomenological potentials, the real part is a WS2 term and the imaginary part is a WS1 
term (pure volume).}
\label{tws1}
\begin{tabular}{cccccc}
\multicolumn{6}{c}{
$^{16}$O$+^{12}$C$~~~$ $R_{V}$=4 fm,  $a_{V}$=1.4 fm}\\\hline
 $E_{\rm lab}$ & $E_{\rm c.m.}$ & V & W & R$_{W}$ & a$_{W}$   \\

[MeV] & [MeV] & [MeV] &  [MeV] & [fm] & [fm]  \\\hline
132   & 56.6 & 293  & 13.4 & 5.900 & 0.603 \\
124   & 53.2 & 290  & 14.1 & 5.712 & 0.636   \\
115.9 & 49.7 & 290  & 13.0 & 5.878 & 0.522   \\
100   & 42.9 & 297  & 10.4 & 6.079 & 0.523  \\
94.8  & 40.6 & 297  & 8.8  & 6.672 & 0.317   \\
80.0  & 34.3 & 297  & 9.0  & 6.557 & 0.322  \\\hline
\multicolumn{6}{c}{
$^{18}$O$+^{12}$C$~~~$ $R_{V}$=4.08 fm,  $a_{V}$=1.38 fm}\\\hline
$E_{\rm lab}$ & $E_{\rm c.m.}$  & V & W & R$_{W}$ & a$_{W}$   \\

[MeV] & [MeV] & [MeV] &  [MeV] & [fm] & [fm]  \\\hline
120 & 48 & 293  & 13.4 & 6.443 & 0.523   \\
100 & 40 & 305  & 13.9 & 6.270 & 0.615   \\
85  & 34 & 324  & 18.3 & 5.930 & 0.562  \\
\end{tabular} 
\end{table}

\begin{table}
\caption{Volume integrals for the potentials obtained  in
$^{16}$O$+^{12}$C and 
$^{18}$O$+^{12}$C. 
$W_{\rm WS2}+W_{D}$ stands for the parameters listed in Table \ref{tws2} and  $W_{\rm WS}$  for the 
parameters in Table \ref{tws1}.}
\label{tintegrali}
\begin{tabular}{c||c|ccc||c|c}
\multicolumn{1}{c||}{$^{16}$O$+^{12}$C}&\multicolumn{4}{c||}{$W_{\rm WS2}+W_{D}$}
&\multicolumn{2}{c}{ $W_{\rm WS}$  } \\ \hline
Energy & J$_{V}$ &  J$_{W}$ &  J$_{WD}$ & J$_{W}$ + J$_{WD}$ &J$_{V}$ &  J$_{W}$  \\ \hline
132   & 312 & 13 & 48 & 61 & 313 & 66 \\
124   & 316 & 24 & 40 & 64 &309 & 64 \\
115.9 & 308 & 29 & 32 & 61 &310&62\\
100   & 308 & 34 & 20 & 54 &317&55\\
94.8  & 304 & 35 & 16 & 51 &317&58\\
80    & 297 & 40 & 14 & 54 &317&56\\\hline
\multicolumn{1}{c||}{$^{18}$O$+^{12}$C}&\multicolumn{4}{c||}{$W_{\rm WS2}+W_{D}$}
&\multicolumn{2}{c}{ $W_{\rm WS}$  } \\ \hline
Energy & J$_{V}$ &  J$_{W}$ &  J$_{WD}$ & J$_{W}$ + J$_{WD}$ &J$_{V}$ &  J$_{W}$  \\ \hline
120 & 296 & 25 & 48 & 73&292&74 \\
100 & 309 & 28 & 47 & 75 &304&73\\
 85 & 325 & 28 & 46 & 74 &323&81\\
\end{tabular} 
\end{table}

\begin{figure}
\caption{Elastic scattering data shown as the ratio to the 
Rutherford cross sections and optical model calculations 
(thick-solid curves) with phenomenological potentials
(left panels with parameters from Table \ref{tws2}, and right panels from Table \ref{tws1}) of the 
$^{16}$O$+^{12}$C at 132, 124, 115.9, and 100 MeV (from top to bottom, left and right, respectively).
The nearside (thin-solid curve) and farside (dashed curve) subamplitudes
for different potentials are plotted.}
\label{elo16}
\end{figure}
\begin{figure}
\caption{Same caption as for Fig. \ref{elo16} but for the $^{18}$O$+^{12}$C elastic scattering at
120, 100, and 85 MeV.}
\label{elo18}
\end{figure} 
\begin{figure}
\caption{Measured elastic angular distributions displayed as ratio to the Rutherford 
scattering of the $^{16}$O$+^{12}$C (stars, panel $a$) and 
$^{18}$O$+^{12}$C scattering (dots, panels $b$ and $c$) at an 
incident energy of 100 MeV. 
The solid curve (parameters from Table \ref{tws2}) and dashed curve 
(parameters from Table \ref{tws1}) 
of panels $a$ and $c$ represent optical-potential fits.
The same fits are presented in Fig. \ref{elo16} (bottom panel)
and Fig. \ref{elo18} (middle panel).
The solid [$P(W_{\rm WS2}+W_{\rm D})$] and dashed curves
[$P(W_{\rm WS})$]  of panel $b$ represent optical-potential calculations
using the parameters obtained in the $^{16}$O$+^{12}$C fit but changing the projectile mass.}
\label{elo16o18}
\end{figure}
\begin{figure}
\caption{The real and imaginary potentials at 100 MeV of the $^{16}$O$+^{12}$C
(left-top panel, and solid curves in bottom panels) and $^{18}$O$+^{12}$C
(right-top panel and dashed curves in bottom panels) for the two
sets of parameters listed in Tables \ref{tws2} and \ref{tws1}.}
\label{poto16o18e100}
\end{figure}
\begin{figure}
\caption{The calculated farside amplitudes for all measured energies
of the $^{16}$O$+^{12}$C system using the parameters from Table \ref{tws2}.
The imaginary strength has been reduced by 50$\%$ to emphasize the
refractive effects.
Curves at different energies have been shifted by a factor
of $100$ for clarity.}
\label{nfo16}
\end{figure}
\begin{figure}
\caption{Same caption as for Fig. \ref{nfo16} but for the $^{18}$O$+^{12}$C system.}
\label{nfo18}
\end{figure}
\begin{figure}
\caption{The position of Airy minima
($\theta_{\rm c.m.}$ versus $E_{\rm c.m.}$) of the $^{16}$O$+^{12}$C system.
The solid curves are results of the $\chi ^{2}$ fit
to the data using the second-order polynomial as the fit function.}
\label{airyo16}
\end{figure}
\begin{figure}
\caption{Same caption as for Fig. \ref{airyo16} but for the $^{18}$O$+^{12}$C system.
}
\label{airyo18}
\end{figure}
\begin{figure}
\caption{Real and imaginary parts of the phenomenological
optical potentials obtained for the $^{12}$C$+^{12}$C, $^{16}$O$+^{16}$O,
$^{16}$O$+^{12}$C, and $^{18}$O$+^{12}$C systems at three
different energies.}
\label{pot-all}
\end{figure}
\end{multicols}

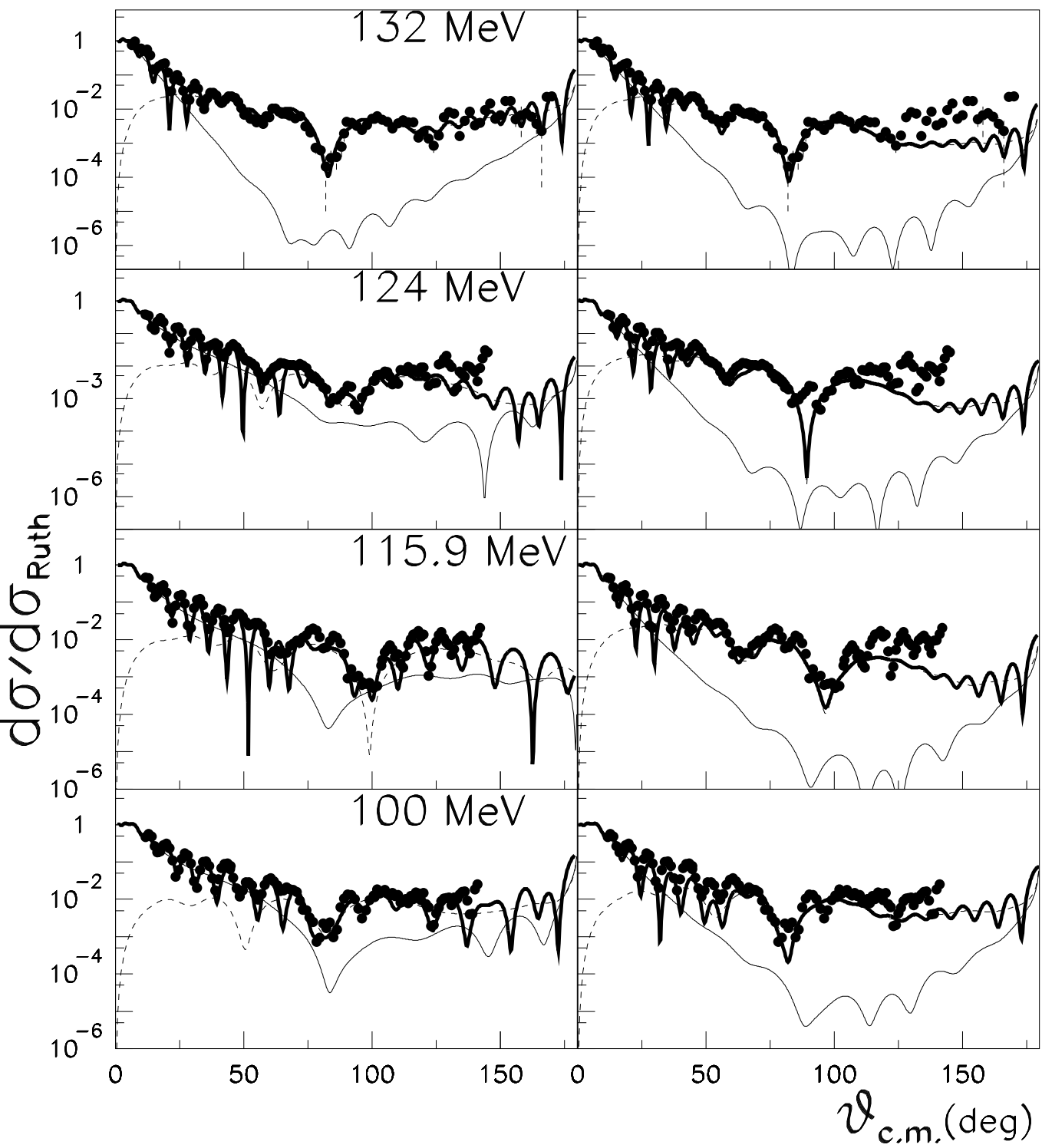
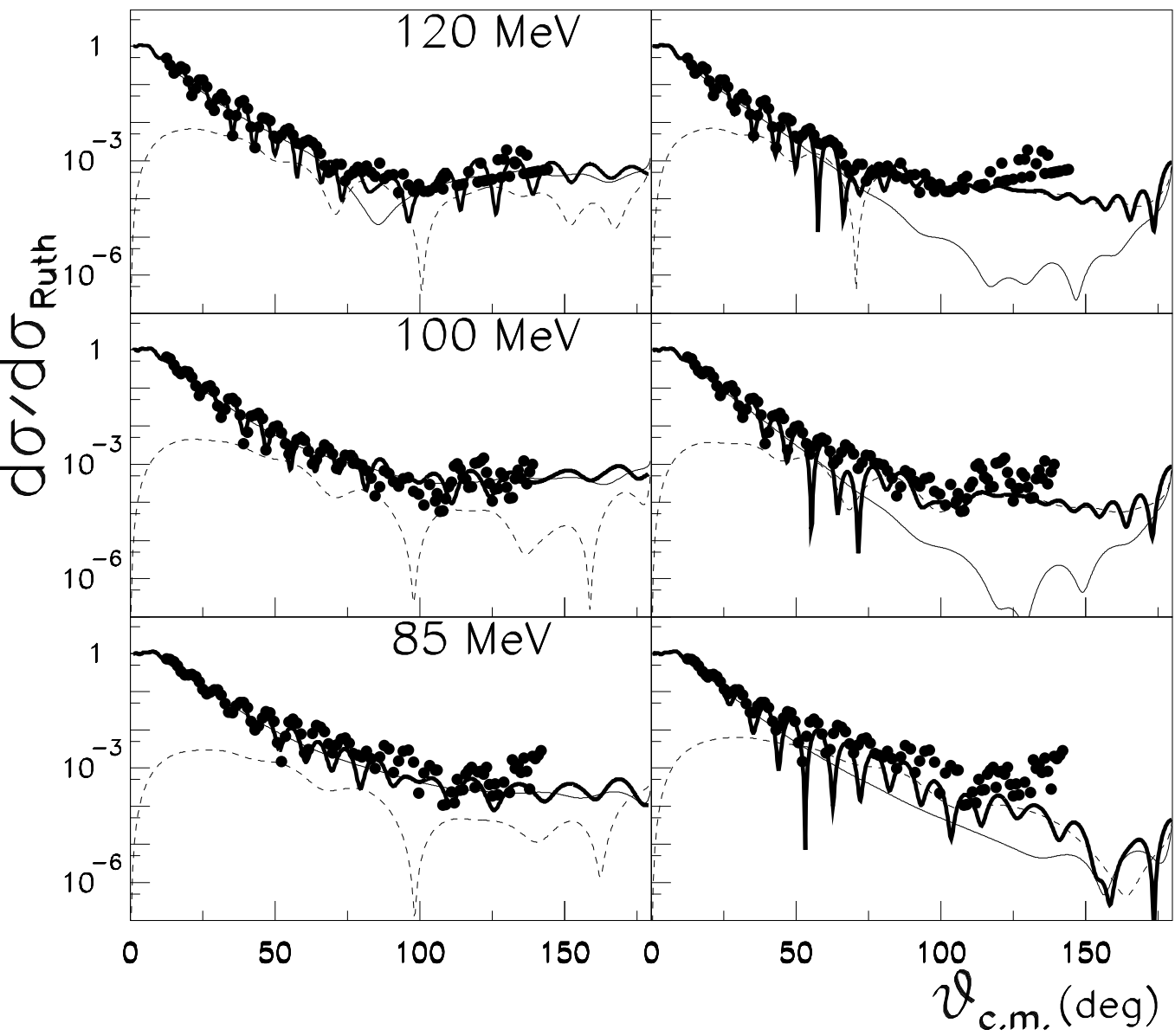
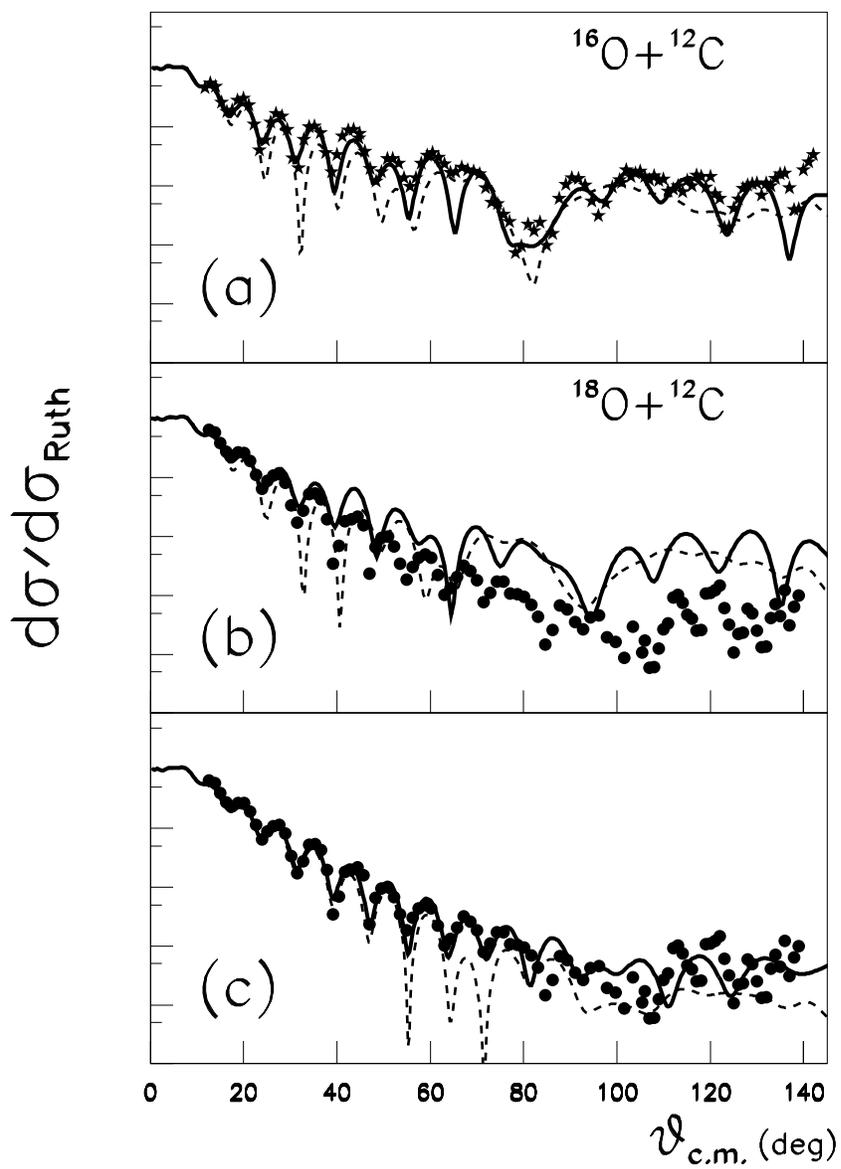
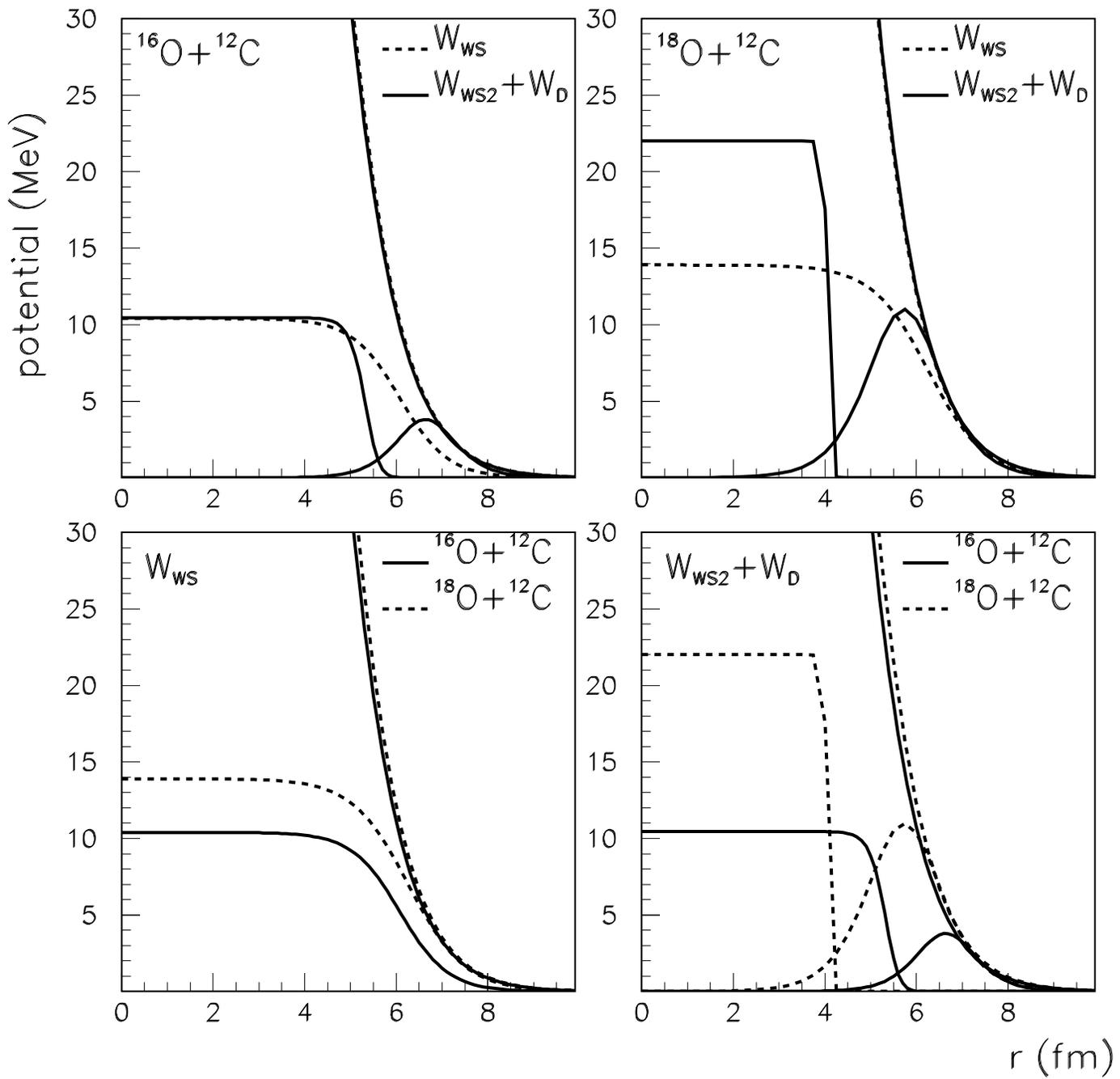
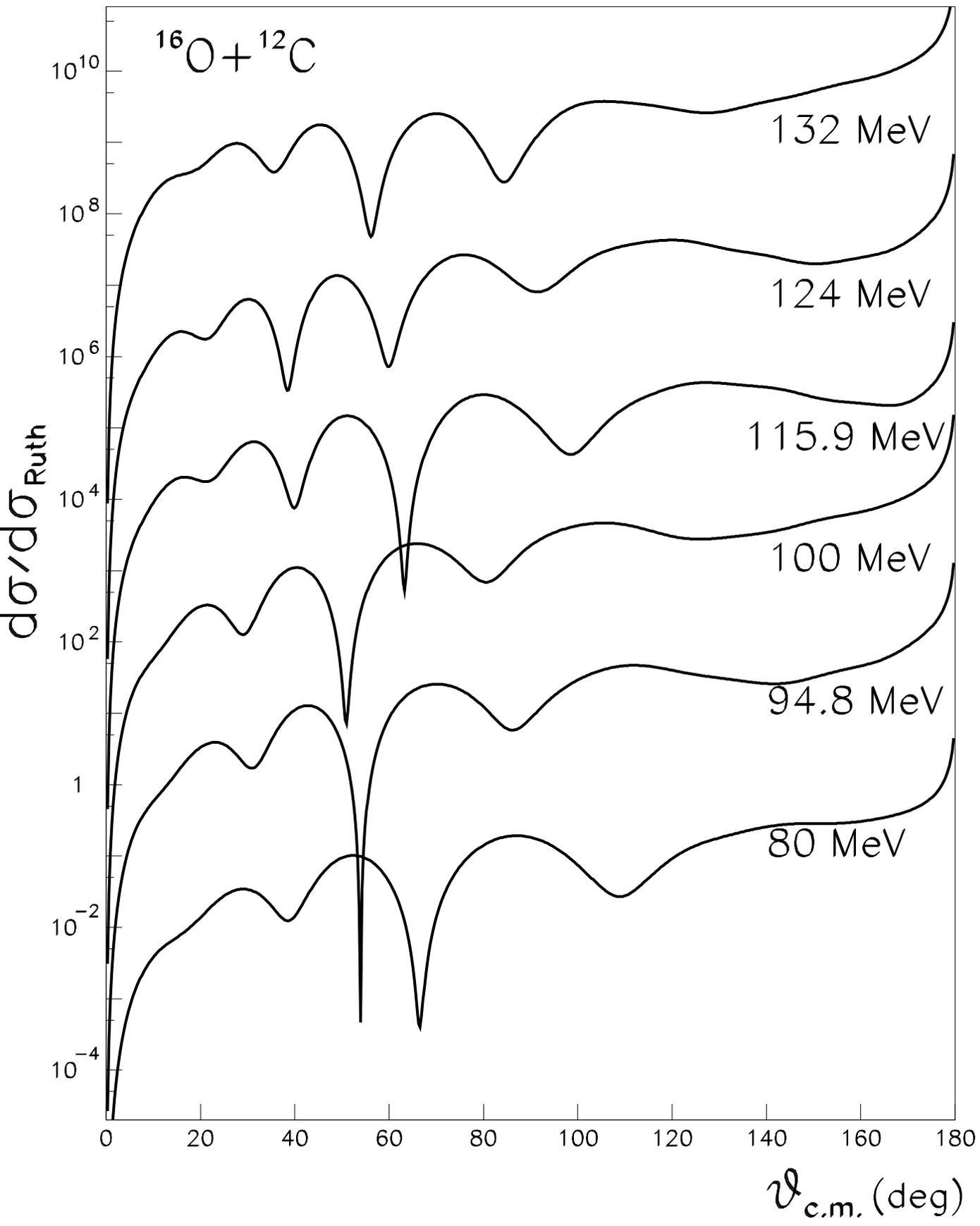
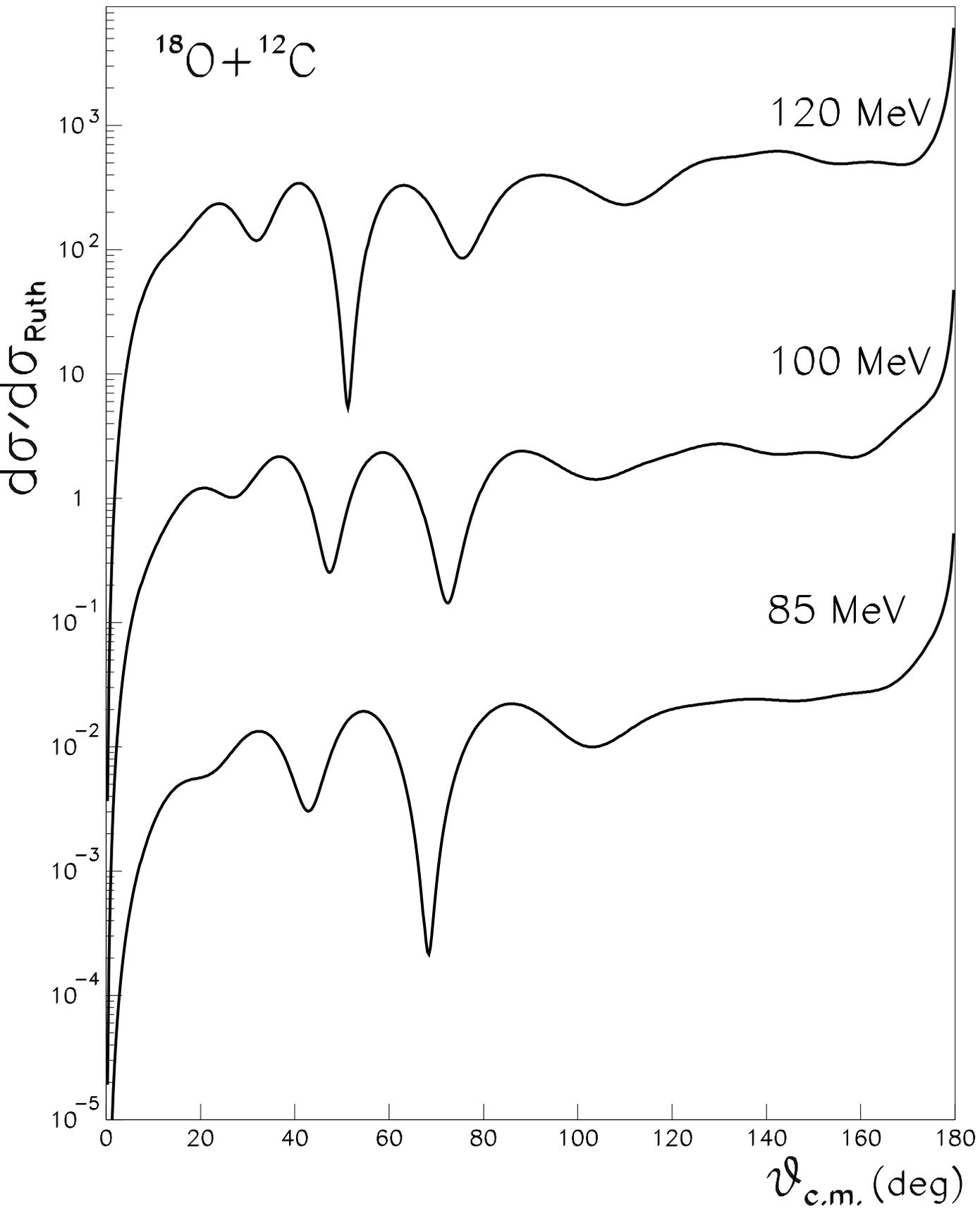
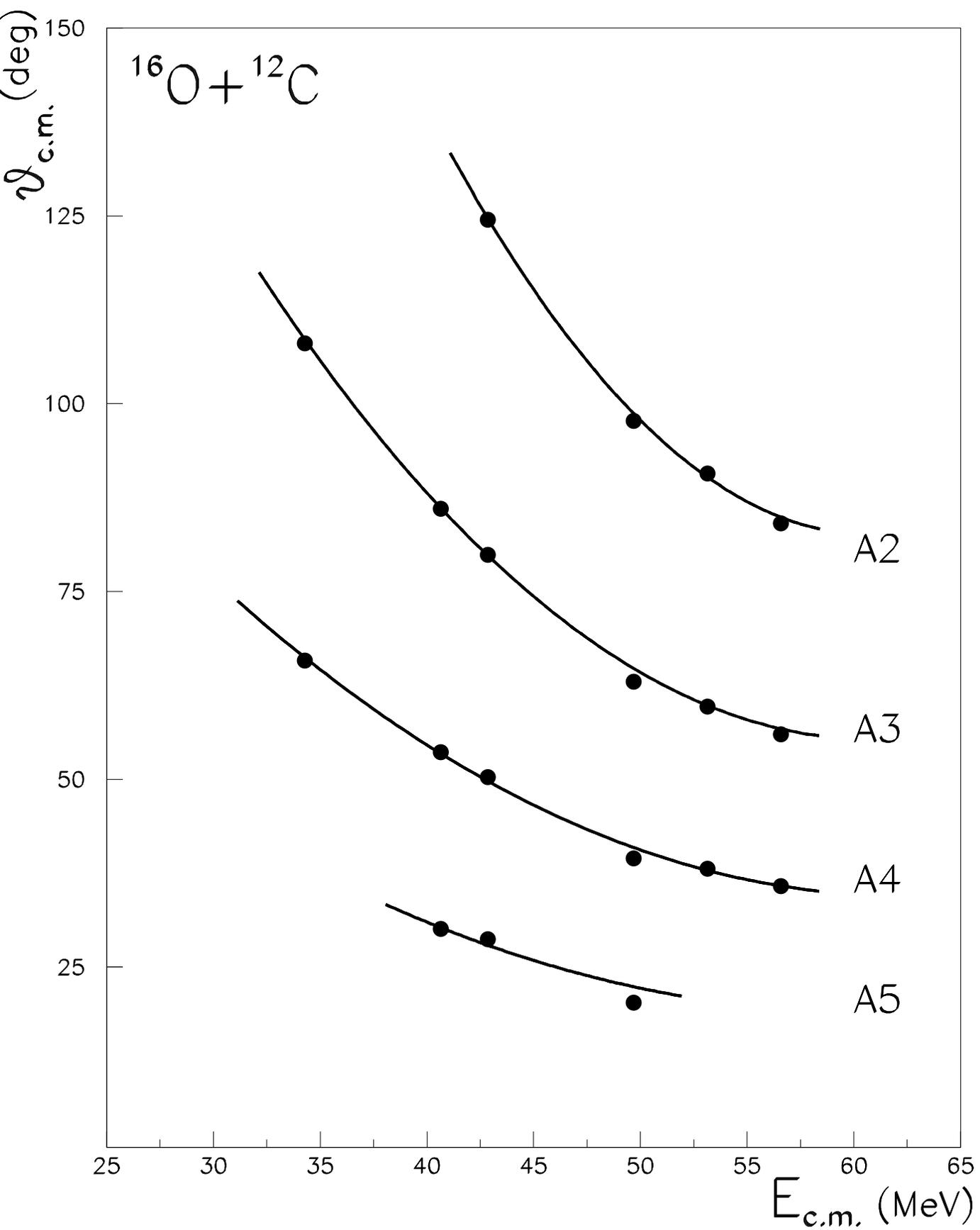
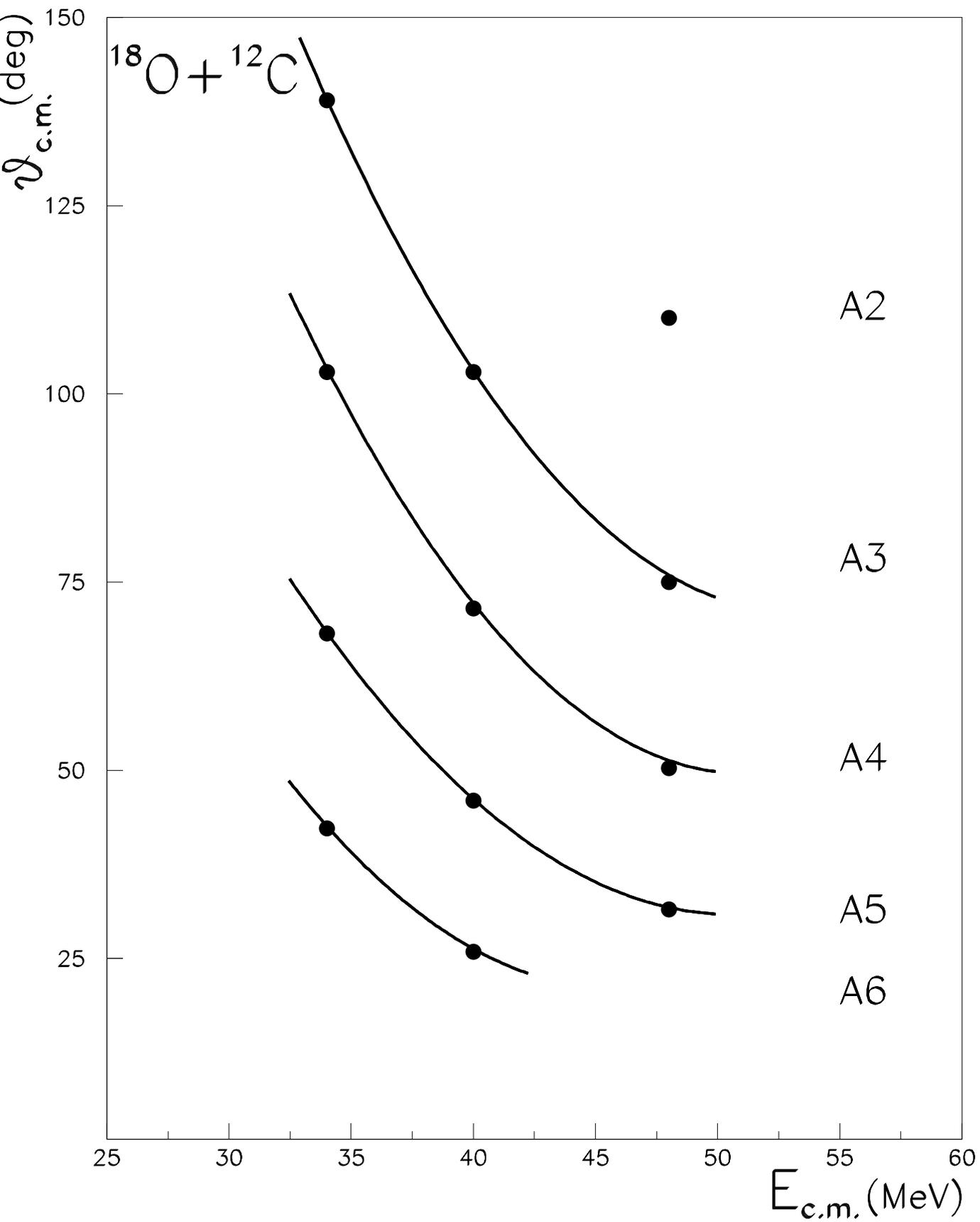
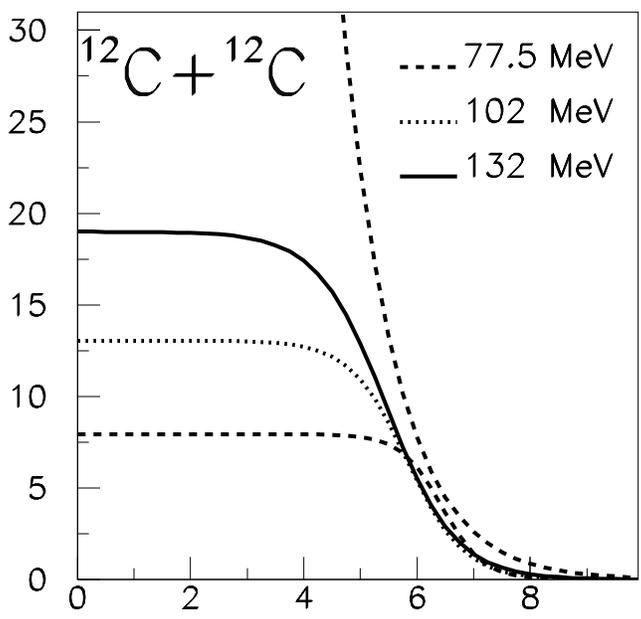
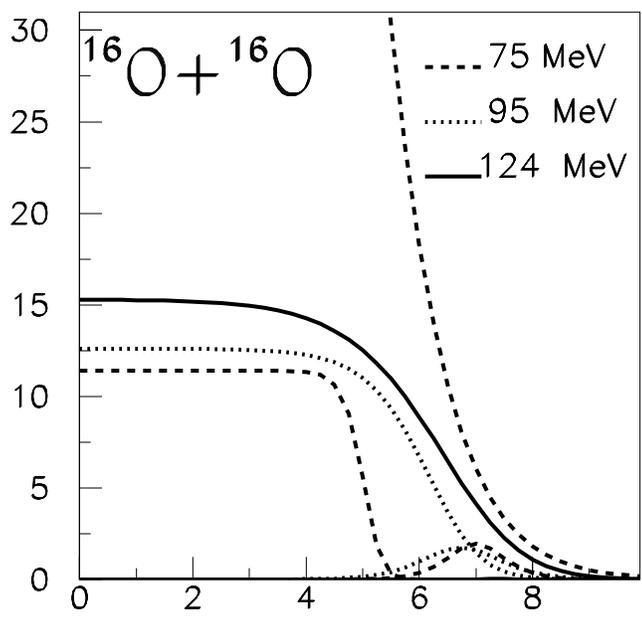
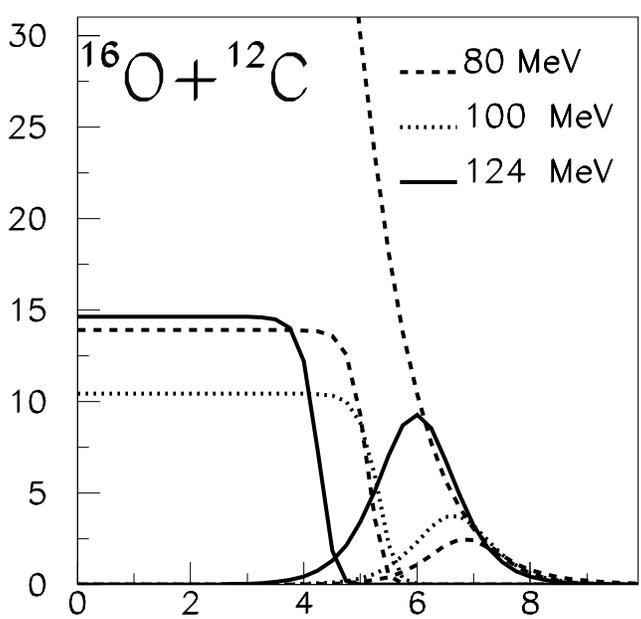
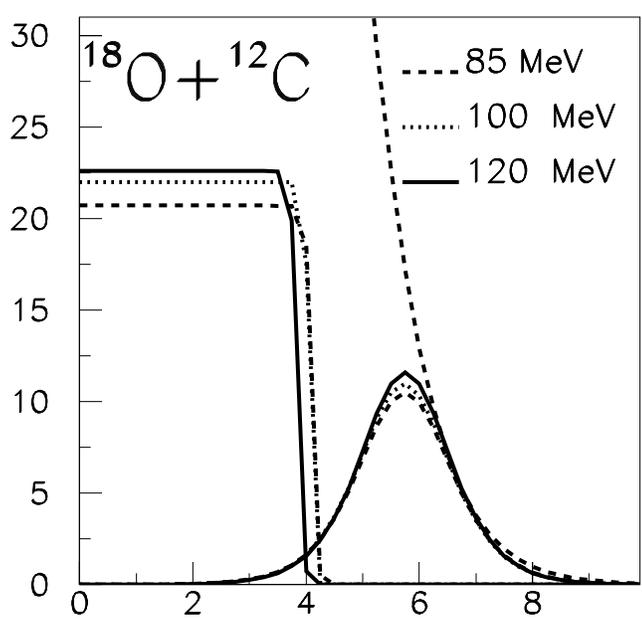
\end{document}